\documentclass[aps,prd,twocolumn,showpacs,groupedaddress]{revtex4}

\usepackage{graphicx}
\usepackage{bm}

\begin{document}

\title{Big Bang Nucleosynthesis: The Strong Nuclear Force meets the Weak Anthropic Principle}

\author{J. MacDonald}
\email[]{jimmacd@udel.edu}
\author{D.J. Mullan}
\affiliation{Department of Physics and Astronomy, University of Delaware, DE 19716}

\date{\today}

\begin{abstract}
Contrary to a common argument that a small increase in the strength of the strong force would lead to destruction of all hydrogen in the big bang 
due to binding of the diproton and the dineutron with a catastrophic impact on life as we know it, we show that provided the increase in strong 
force coupling constant is less than about 50\% substantial amounts of hydrogen remain. The reason is that an increase in strong force strength 
leads to tighter 
binding of the deuteron, permitting nucleosynthesis to occur earlier in the big bang at higher temperature than in the standard big bang. 
Photodestruction of the less tightly bound diproton and dineutron delays their production to after the bulk of nucleosynthesis is complete. The 
decay of the diproton can, however, lead to relatively large abundances of deuterium.
\end{abstract}


\maketitle

\section{Introduction} 

The weak anthropic principle has often been used to infer limits on the range of the strong force strength or coupling parameters consistent with 
life as we know it. A common argument is that a small increase of the strong force strength will bind the dineutron and the diproton, leading to 
a large increase in the rate of the p+p and n+n reactions, so that big-bang nucleosynthesis (BBN) leads to all the protons being converted into 
isotopes of helium, leaving no hydrogen necessary for chemistry vital to life. For example, Dyson \cite{1971SciAm.225...50D} writes "If a 
helium-2 nucleus could exist, the proton  proton reaction would yield a
 helium-2 nucleus plus a photon, and the helium-2 nucleus would in turn spontaneously decay into a deuteron, a positron and a neutrino. As a 
consequence there would be no weak interaction hang-up, and essentially all of the hydrogen existing in the universe would have been burned to 
helium even before the first galaxies had started to condense." Barrow and Tipler \cite{1986acp..book.....B} state on p322 of their book "If the 
strong interaction were a little stronger the diproton would be a stable bound state with catastrophic consequences  all the hydrogen in the 
Universe would have been burnt to He2 during the early stages of the Big Bang and no hydrogen compounds or long-lived stable stars would exist 
today. If the diproton existed we would not!"

In this paper, we consider how BBN is altered by the existence of bound diproton and dineutron nuclei, taking into account some physical 
processes that so far have been overlooked. To relate the binding energies of the diproton, the dineutron and the deuteron to the relative 
strength of strong force, we use the same square well potential model as Barrow \cite{1987PhRvD..35.1805B}, who found that a 9\% increase in the 
strong force coupling constant, $\alpha_s$, is sufficient to bind the dineutron and a 13\% increase will bind the diproton. The needed increase 
in coupling constant to bind the diproton was confirmed by Pochet et al \cite{1991A&A...243....1P} for a more realistic nuclear potential. For 
small increases in $\alpha_s$, the diproton and dineutron binding energies are sufficiently small that photodestruction will prevent buildup of
large amounts of these isotopes before freeze out occurs. 

In section 2, we briefly review the physics of BBN relevant to our investigation of the effects of increased $\alpha_s$. In section 3 we
describe how we determine the needed rates for reactions involving dineutrons and diprotons. Due to the electrostatic repulsion of the
protons, there is a narrow range of strong force strength for which the dineutron is bound but the diproton is not. Nucleosynthesis in this
regime is considered in section 4.  In section 5, we consider the regime in which the diproton is also bound.  Finally, in section 6 we
give our conclusions.

\section{BBN in the radiation dominated era}

In the standard hot big bang \cite{1973ARA&A..11..155H}, nucleosynthesis takes place in the radiation dominated era that begins after
electron pair annihilation is complete. At the beginning of this era the radiation and matter temperature is $\sim 4 ~ 10^9$ K. Although the 
$p + n \rightarrow d + \gamma$ reaction is rapid, due to the relatively small binding energy of the deuteron, photodestruction prevents
significant amounts of $^2$H being
formed until the temperature has dropped to $\sim 10^9$ K. Further reactions continue, building up an appreciable amount of $^4$He and
traces of other light elements, until freeze out occurs when the Universe is about 15 minutes old and has a temperature of  $\sim 5 ~ 10^8$ K.
During the radiation era, the temperature varies with time, $t$, as
\begin{equation}
		T_9 =13.8~ t^{-1/2}
\end{equation}
where $T_9$ is the temperature in units of $10^9$ K and $t$ is measured in s. The baryonic density in g cm$^-3$ is approximately given by
\begin{equation}
		\rho = 3.3 ~ 10^4 \eta T_9^3
\end{equation}
where $\eta$ is the ratio of the number of baryons to the number of photons.
To explore the effects of bound diproton and dineutron on BBN, we have written a computer program that includes all the important reactions
 \cite{1966ApJ...146..542P} \cite{1967ApJ...148....3W} plus a few more relevant to universes in which the diproton and dineutron are stable. 
Because the binding energy of the deuteron is increased when $\alpha_s$ is increased, nucleosynthesis can begin at higher temperatures than for 
standard BBN. We have modified equations (1) and (2) to include conditions in the leptonic era. We take $T_9 = 50$ as the initial temperature.  
The initial abundances of neutrons and protons are set to their equilibrium values, with all other abundances set to zero. The value of $\eta$ is 
taken to be 4 10$^{-10}$, which gives for the standard big bang good agreement with observed light element abundances. As a test of our code, we 
have compared our standard big bang nucleosynthesis results with those obtained with the public\_bigbang code  (which can be downloaded from
 http://cococubed.asu.edu/code\_pages/net\_bigbang.shtml).

\section{Adopted rates for reactions for reactions involving dineutrons and diprotons}

The production of diprotons and dineutrons in the big bang will depend on the competition for neutrons and protons between the  $p + p 
\rightarrow pp + \gamma$ and $n + n \rightarrow nn + \gamma$ reactions and the $p + n \rightarrow d + \gamma$  reaction. To determine the rates 
of the $p + p \rightarrow pp + \gamma$ and $n + n \rightarrow nn + \gamma$ reactions, we make use of known results for the $p + p \rightarrow d + 
e^+ + \nu$ reaction.  We assume that the intrinsically nuclear part of the interaction potential is the same for all three reactions. To account 
for the difference in rates of the $p + p \rightarrow d + e^+ + \nu$ reaction in which the intermediate state experiences a weak decay and the 
two reactions in which the intermediate state decays electromagnetically, we introduce a factor $f_{w-e}$. Using the $p+p$ S-factor 
\cite{1999NuPhA.656....3A}, a straightforward integration for the $n + n \rightarrow nn + \gamma$ reaction gives for the reaction rate per 
particle in cm$^3$ mol$^{-1}$ s$^{-1}$
 \begin{equation}
		N_A \langle \sigma v \rangle = f_{w-e} 10^{-15}(1.78 + 1.80 T_9 + 1.99 T_9^2)/T_9^{1/2}
 \end{equation}
For the rate of the $p + p \rightarrow pp + \gamma$ reaction, we simply multiply the $p + p \rightarrow d + e^+ + \nu$ rate by $f_{w-e}$,
 \begin{eqnarray}
		&& N_A \langle \sigma v \rangle =f_{w-e} 10^{-15}(4.08 + 15.6 T_9 + 6.16 T_9^2 + 0.588 T_9^3\nonumber\\
		&&- 0.0465 T_9^4)e^{-3.381/T_9^{1/3}}/T_9^{3/2}
 \end{eqnarray}

At first sight it might be thought that $f_{w-e}$ could simply be obtained from consideration of the $p + n \rightarrow d + \gamma$ reaction.
However the comparison is complicated by the effects of nucleon spin on the nuclear interaction. The stability of the deuteron is a result of the 
nuclear force being stronger when the nucleons have parallel spin than when the spins are opposite. The Pauli exclusion principle requires that 
the nucleons in the diproton and dineutron have opposite spin. We can roughly estimate $f_{w-e}$ by considering the nuclear S-factors for similar 
reactions.  For example, comparing $S_0 = 2.5 ~ 10^{-4}$ KeV barn for d(p,$\gamma$)$^3$He with $S_0 = 3.8 ~ 10^{-22}$ KeV barn for p(p,
$\beta^+\nu$)d \cite{1968psen.book.....C} indicates that $f_{w-e} \sim 10^{18}$. The nuclei in this reaction do not have the same spins as the 
nuclei in the reactions of interest. However a similar result is obtained by considering a reaction with the correct spins, 
$^{13}$N$ + p \rightarrow ^{14}$O$ + \gamma$. The non-resonant contribution to this reaction has 
$S_0 \sim 3 ~ 10^{-4}$ KeV barn\cite{1999NuPhA.656....3A}. 
In figure 1, we plot the ratios ${\langle \sigma v \rangle}_{np}/ {\langle \sigma v \rangle}_{nn}$ and ${\langle 
\sigma v \rangle}_{np}/{\langle \sigma v \rangle}_{pp}$ against $T_9$ for the case in which a value of $10^{18}$ is adopted for $f_{w-e}$. For the $p + n \rightarrow d + \gamma$ reaction rate, we have used \cite{2006PhRvC..74b5809A}
 \begin{widetext}
 \begin{eqnarray}
      &&N_A \langle \sigma v \rangle = 4.42~10^4(1.0+3.75 T_9+1.93 T_9^2+0.747 T_9^3+0.0197 T_9^4+3.00491~10^{-6}T_9^5)/\nonumber\\
      &&(1.0+5.47 T_9+5.62 T_9^2+0.489 T_9^3+7.47~10^{-3} T_9^4)
 \end{eqnarray}
 \end{widetext}
We see that at temperatures relevant to BBN, dineutron and diproton production will not be important unless $f_{w-e} \gtrsim 10^{18}$. Because of the uncertainties involved in estimating $f_{w-e}$, in the following we consider a range of $f_{w-e}$ values.

\begin{figure}
\includegraphics[scale=0.5]{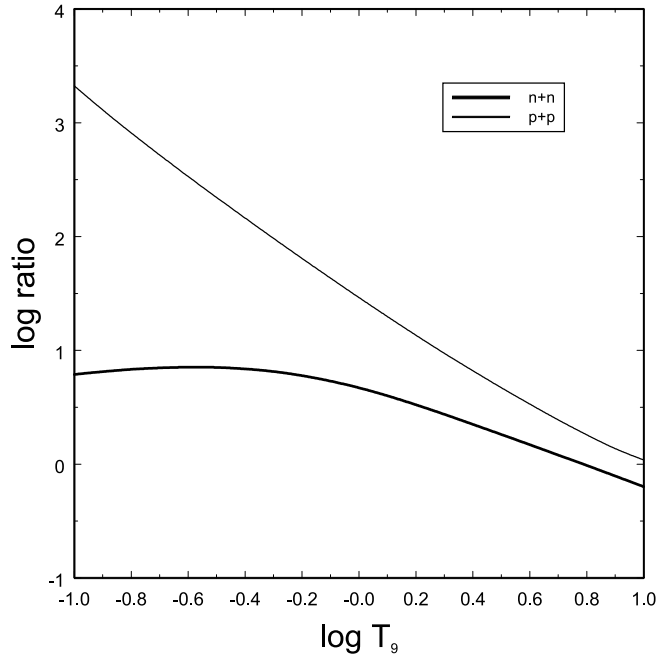}
\caption{\label{fig_1} Dependence of the ratios $ {\langle \sigma v \rangle}_{np}/ {\langle\sigma v \rangle}_{nn}$ and ${\langle \sigma v
 \rangle}_{np}/{\langle \sigma v \rangle}_{pp}$ on temperature in units of $10^9$ K for $f_{w-e} = 10^{18}$}
\end{figure}

The dineutron and diproton production will also depend on the rates of the reverse reactions $pp + \gamma \rightarrow p + p $  and $nn + \gamma
 \rightarrow n + n $. For small increases in the strength of the strong force coupling constant, the binding energies of dineutron and diproton 
will be of order $kT$ during the nucleosynthesis phase of the big bang. Hence, the threshold value of $e^{E_\gamma/kt}$, where $E_\gamma$ is the 
photon energy, is not large compared to unity, and the usual approximation that the reverse rate is the forward rate multiplied by a factor
\begin{equation}
  \frac{\lambda_\gamma}{N_A \langle \sigma v \rangle} = 7.07 ~ 10^9  ~ T_9^{3/2} ~  e^{-Q/kT}
\end{equation}
cannot be used indiscriminately. Although the rate of the $nn + \gamma \rightarrow n + n $ reaction can be evaluated analytically in terms of 
Debye functions, we find it simpler to evaluate this rate numerically, together with the $pp + \gamma \rightarrow p +p$ reaction rate. For a 
binding energy $Q = 150$ keV, we find that the above approximation underestimates the reverse rate by only 13\% at $T_9 = 4$. We determine a 
correction to the approximate rate by multiplying it by a factor of form $1/(1+a_1 q + a_2 q^2 + a_3 q^3)$  where $q = e^{-Q/kT}$. The 
coefficients $a_1$, $a_2$, $a_3$ are found by fitting to the numerical results. An important consequence of an increase in $\alpha_s$ is that the 
binding energy of the deuteron will also be increased. This will reduce its photodestruction rate and allow $^2$H production to occur at higher 
temperature than in the standard big bang.  We take this increase in the deuteron binding energy, $Q_d$ into account in calculating the
 photodestruction rate of $^2$H.

We determine the binding energies, $Q_d$, $Q_{nn}$ and $Q_{pp}$ by using a square well potential for the nucleon interaction of depth $V$ and  
radius $r$. To characterize the strength of the strong force, we use the relative strong charge, $G$, which is related to the strong force
 coupling constant by
\begin{equation}
    \alpha_s = G^2 \frac{g_0^2}{\hbar c}
\end{equation}
where $g_0$ is the standard strong charge value. The depth of the potential is then
\begin{equation}
	V = G^2 V_0
\end{equation}
We use the same values for $V_0$ and $r$ as Barrow \cite{1987PhRvD..35.1805B}. For the deuteron $J^\pi = 1^+$ ground state $V_0$ = 36.2 MeV and 
$r$ = 2.02 fm. For the $J^\pi = 0^+$ $^2$H and dineutron states, $V_0$ = 14.0 MeV and $r$ = 2.59 fm. The diproton potential is taken to be 0.56 
MeV shallower than the dineutron potential to account for the Coulomb energy. The dependences of the resulting binding energies on $G$ are shown
 in figure 2.

\begin{figure}
\includegraphics[scale=0.5]{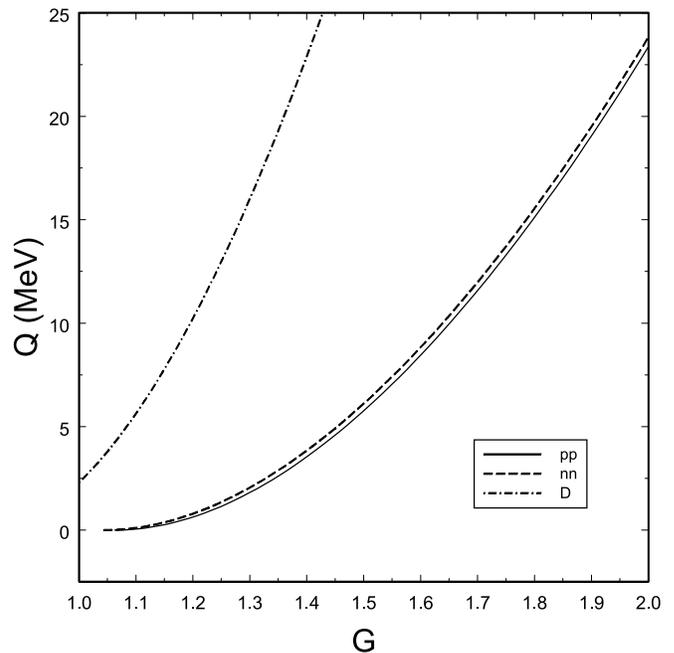}
\caption{\label{fig_2} Dependence of the deuteron, dineutron and diproton binding energies in MeV on the relative strong charge $G$.}
\end{figure}

We now consider the rates of leptonic transformations between dineutrons, diprotons and deuterons. For small binding energies, we expect the 
time for dineutron decay to be comparable to that of the neutron. Similarly, since the proton is stable, the lifetime of the diproton is likely 
to be quite long. A comparison with the leptonic rates between n and p \cite{1994ADNDT..56..231O} indicate that the enhancements of the overall 
rates, including electron capture, at BBN temperatures will be modest and hence for simplicity we neglect these enhancements. To estimate the 
weak rates, we use the $ft$ factors for the corresponding transitions in the beta decays of the analog nuclei $^{14}$O and $^{14}$C. 

The major $^{14}$O decay channel is to an excited $J^\pi = 0^+$ state 2.313 Mev above the ground state of $^{14}$N (99.3\%, log $ft$ = 3.4825). 
This is analogous to diproton decay to the spin singlet state of $^2$H. There are also decays to the $1^+$ ground state (log $ft$ = 7.279) and to 
a $1^+$ excited state at 3.948 MeV (log $ft$ = 3.131).  We have calculated the Fermi integrals with the relativistic form of the Fermi factor for 
the two higher energy decay channels which correspond to those of the diproton. For the diproton decay channel corresponding to the dominant 
$^{14}$O decay, we find that an accurate approximation to the decay time scale is
\begin{equation}
	\log t_{1/2} = 3.490 - 5 \log E_+
\end{equation}
where the maximum positron kinetic energy measured in MeV is, in terms of binding energies,
\begin{equation}
	E_+ = Q_{nn} - Q_{pp} - (m_n - m_p) -m_e = Q_{nn} - Q_{pp} - 1.8043.
\end{equation}
For the decay to the ground state
\begin{equation}
	\log t_{1/2} = 7.287 - 5 \log E_+
\end{equation}
where now
\begin{equation}
  E_+ = Q_D - Q_{pp} - (m_n - m_p) -m_e = Q_D - Q_{pp} - 1.8043.
\end{equation}
	
The analog of the dineutron, $^{14}$C, decays only to the ground state of $^{14}$N with half-life $t_{1/2} = 1.8~10^{11}$ s (log $ft$ = 9.040). 
We find that the half-life for the corresponding dineutron decay is approximately related to the maximum electron kinetic energy in MeV by
\begin{equation}
	t_{1/2} = (63/E_-)^5
\end{equation}
where
\begin{equation}
	E_- = Q_D - Q_{nn} - (m_n - m_p - m_e) = Q_D - Q_{nn} - 0.7823.
\end{equation}
Since we expect the dineutron will also have a decay channel to the $0^+$ excited state of $^2$H (which will exist if the dineutron is bound), we 
use for this channel the $ft$ factor for the corresponding transition in $^{14}$O.  The energy difference between the dineutron and the $^2$H 
singlet state will always be about the difference in mass of the neutron and proton, ~1.3 MeV, and the $f$-factor will then be about 1.8, which 
gives $t_{1/2} = 1.5~10^3$ s. Combining the rates for the two decay channels gives
\begin{equation}
	\lambda_{nn} = 4.62~ 10^{-4} +(E_-/58.5)^5.
\end{equation}
The resulting decay time scales, $\lambda^{-1}$, are plotted against $G$ in figure 3.

\begin{figure}
\includegraphics[scale=0.5]{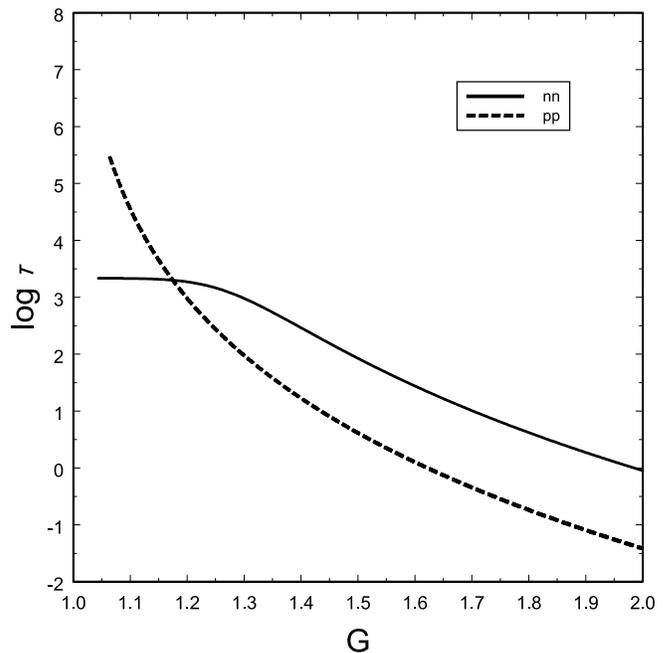}
\caption{\label{fig_3} Dependence of the dineutron and diproton beta decay life times in seconds on the relative strong charge $G$.}
\end{figure}

Finally, we need to also consider additional reactions that arise when the diproton and dineutron are bound.  The most rapid reactions are likely 
to be $pp + n \rightarrow d + p$ and $nn + p \rightarrow d + n$.  Due to the complexity of calculating reaction rates even for few nucleon 
systems (see for example Marcucci et al.  \cite{2006NuPhA.777..111M}), we settle for estimating when these two reactions are likely to be 
important by comparing the results of calculations in which these reactions are completely neglected with the results of calculations in which
 they are assumed to be instantaneous. 

\section{The \lowercase{n+n $\rightarrow$ nn} reaction regime}

Due to the electrostatic repulsion of the protons, there is a narrow range of strong force strength for which the dineutron is bound but the 
diproton is not. According to the square well potential model, this range is $1.043 < G < 1.063$. The dineutron binding energy in this regime is 
$Q_{nn}  = 0 - 14$ KeV. The relevant new reactions for BBN are $n +n \rightarrow nn + \gamma$, $nn + \gamma \rightarrow n + n$  and $nn 
\rightarrow d + e^- + \bar{\nu}$. In figure 4, we show for $G = 1.06$ how the final mass fractions of $^1$H, $^4$He and the dineutron depend on  
$f_{w-e}$ when only the $n + n \rightarrow nn + \gamma$ reaction is included.  We see that there is no significant production of dineutrons 
unless $f_{w-e} \gtrsim 10^{15}$.  Also the H abundance increases with $f_{w-e}$ because the $n +n \rightarrow nn + \gamma$ reaction removes the 
neutrons before the $n + p \rightarrow d + e^+ +\nu$ reaction can take place.

\begin{figure}
\includegraphics[scale=0.5]{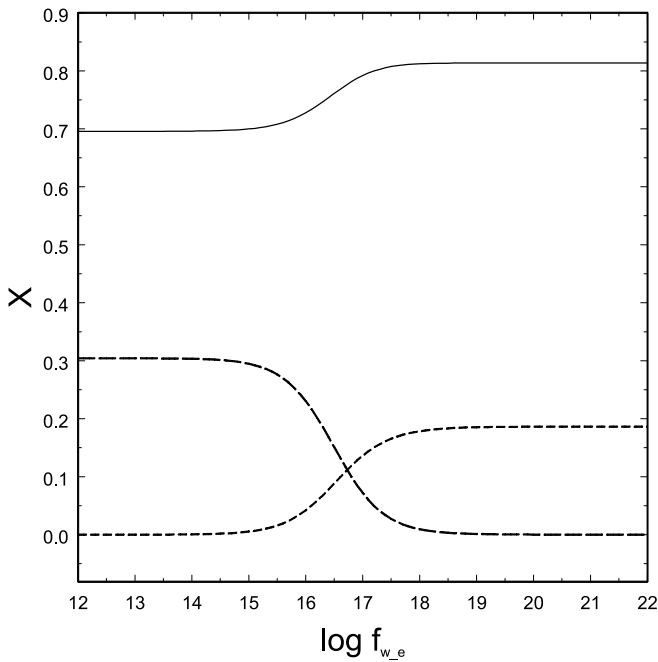}
\caption{\label{fig_4} Final mass fractions of $^1$H (solid line), $^4$He (long dash line) and dineutron (short dash line) when only the n+n 
$\rightarrow$ nn reaction is included.}
\end{figure}

In figure 5, we show the final dineutron abundance when the $nn + \gamma \rightarrow n + n$ reaction isalso included for different values of $G$. 
Dineutron production is small unless its binding energy is comparable to that of the deuteron.  Otherwise, the neutrons are removed by proton 
capture before significant amounts of dineutron can be produced. For values of $G$ large enough to give the dineutron a binding energy greater 
than 2 MeV, the diproton would certainly be bound.

\begin{figure}
\includegraphics[scale=0.5]{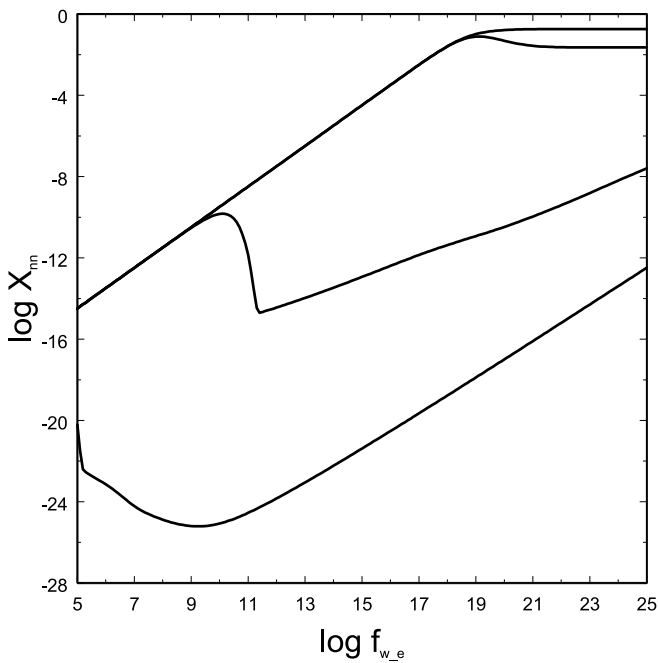}
\caption{\label{fig_5} Final abundance of dineutron for $G$ values from bottom to top of 1.25, 1.50, 1.75 and 2.00.}
\end{figure} 

The $n + n \rightarrow nn + \gamma$ reaction could lead to significant production of $^2$H if the dineutron leptonic decay occurs more quickly 
than its photodestruction. In figure 6, we show how the final mass fraction of $^1$H depends on the time scale of the $nn \rightarrow d + e^- + 
\bar{\nu}$ reaction for $G = 1.06$. In order to have significant dineutron production to occur, we have set $f_{w-e} = 10^{20}$.  We see that, if 
fast enough, the leptonic decay increases the hydrogen abundance. This is because the set of reactions 
	$$n + n \rightarrow nn + \gamma$$
	$$nn \rightarrow d + e^- + \bar{\nu}$$
	$$d + \gamma \rightarrow n + p$$
converts a neutron into a proton. For this to happen, the dineutron decay must occur on a time scale of $10^{-12}$ s or less, which is much less 
than the estimate above, $\tau_{nn} \sim 10^3$ s. Hence it is unlikely that this set of reactions is important.

Finally we consider the effects of the $nn + p \rightarrow d + n$ reaction.  To gauge the importance of this reaction, we assume that it is 
instantaneous. The final mass fractions of $^1$H, $^4$He are shown in figure 7, again for $G = 1.06$. We see that this reaction leads to small 
reductions in the H abundance for $f_{w-e}  > 10^{19}$. Hence this reaction does not have a major effect on BBN when only the dineutron is bound.

\begin{figure}
\includegraphics[scale=0.5]{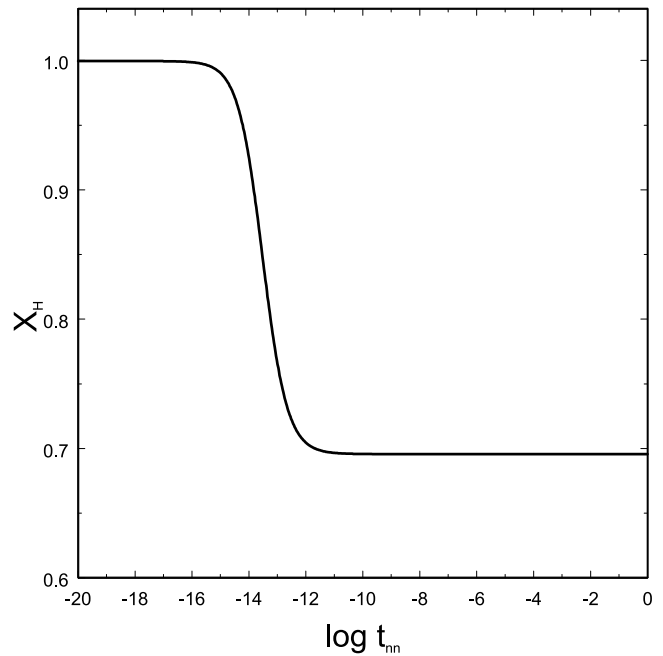}
\caption{\label{fig_6}  Dependence of hydrogen final abundance for $G =1.06$ on dineutron life time (in s).}
\end{figure} 

To summarize the results presented in this section, we find that for values of the strong force coupling constant at which the dineutron is bound 
and the diproton is unbound, there are no catastrophic impacts on BBN.
 
\begin{figure}
\includegraphics[scale=0.5]{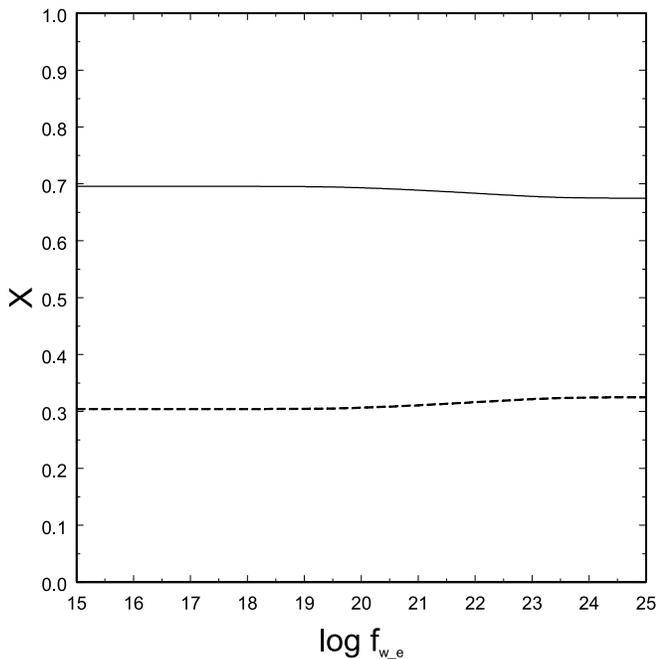}
\caption{\label{fig_7}  $^1$H and $^4$He mass fractions when the $nn + p \rightarrow d + n$ reaction is included.}
\end{figure}

\section{The \lowercase{p+p $\rightarrow$ pp} regime}

For $G > 1.065$ both the diproton and the dineutron are bound. Increased $G$ also binds the deuteron more tightly, allowing it to be formed 
earlier in the big bang at higher temperatures, where the less tightly bound diproton and dineutron are easily destroyed by energetic photons. We 
first consider only the effects of increased $G$ on deuteron binding by setting $f_{w-e} = 0$.  Figure 8 shows how the final hydrogen and helium 
abundances depend on $G$. In the standard big bang $^2$H production begins when the temperature has dropped to about $10^9$ K. For $G \gtrsim 
1.2$, $Q_d$ is high enough that $^2$H production begins in the leptonic era. The final $^1$H abundance is then approximately the difference in 
the equilibrium proton and neutron abundances at the temperature at which photodestruction of $^2$H becomes unimportant. 

\begin{figure}
\includegraphics[scale=0.5]{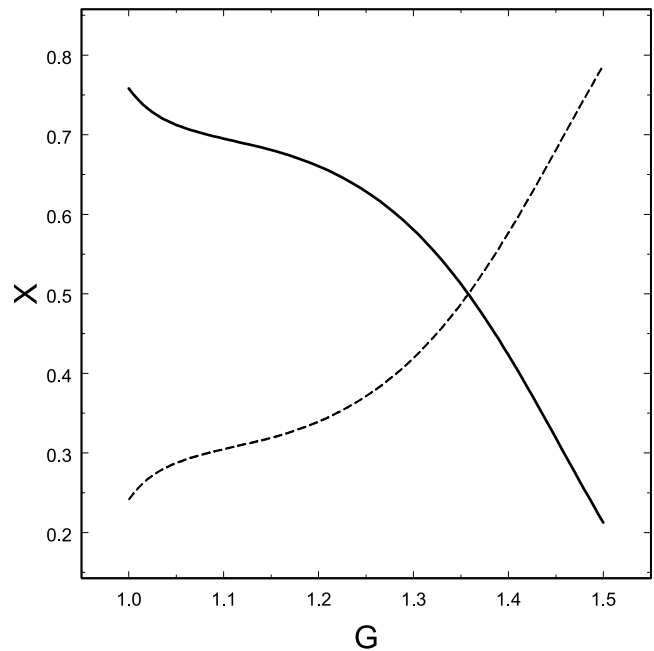}
\caption{\label{fig_8}  Dependence of the final abundances of H (solid line) and $^4$He (broken line) on $G$ in the absence of production of 
dineutrons and diprotons.}
\end{figure} 
 
When $f_{w-e} > 0$, if diproton production occurs it does so long after the primordial neutrons have been consumed in the reactions that lead to 
$^4$He. Hence essentially no dineutrons are produced. The amount of additional $^4$He produced depends on the temperature at which diproton 
production occurs. If the diproton is lightly bound the temperature will be too low for further nuclear processing except for the decay to $^2$H.  
If the diproton is tightly bound the temperature can be high enough for further nuclear processing to $^4$He. In either case diproton production 
does further reduce the hydrogen abundance. The second phase of $^4$He production occurs only if
 \begin{equation}
	f_{w-e} \gtrsim \frac{10^{14}}{\left(  G ~ - ~ 1.113 \right) ^{13/3}}
\end{equation}
 A typical situation is shown in figure 9. Here $G = 1.3$ and $f_{w-e} = 10^{18}$. Initially weak interactions convert neutrons to protons. When 
the Universe is 2 s old the temperature is $8 ~  10^9$ K, which for $Q_d$ = 16 MeV is low enough for $^2$H production to occur. The $^2$H is 
quickly converted to $^4$He, so that by age 10 s, this initial phase of nucleosynthesis has finished. A second phase of nucleosynthesis occurs at   
age 500 s, when the temperature, $T= 6 ~ 10^8$ K, is low enough for production of diprotons, which have a binding energy of 1.8 MeV.  The beta 
decay life time of the diproton is ~ 100 s, and hence diprotons decay to $^2$H before significant cooling by expansion occurs. The temperature is 
sufficiently high that the $^2$H is converted to $^4$He. If the diproton binding energy was lower, then diproton production would occur at lower 
temperature and only $^2$H would be made in the second phase of nucleosynthesis. Note the small amount of neutrons released in the second phase 
of nucleosynthesis. These are produced by the reaction sequence $d  + d \rightarrow t + p$, followed by $t + t \rightarrow $$^4$$He + n + n$.

\begin{figure}
\includegraphics[scale=0.5]{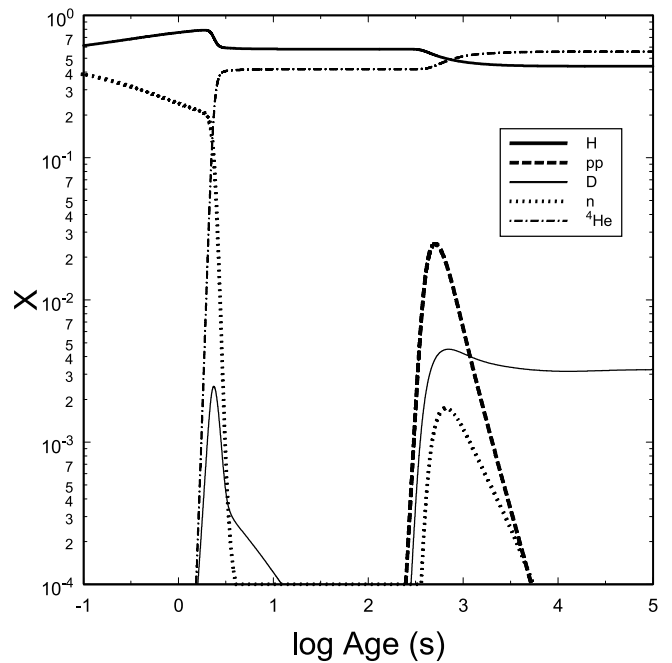}
\caption{\label{fig_9} Final mass fractions of n, $^1$H, $^2$H, $^2$He and $^4$He for $G = 1.3$ and $f_{w-e} = 10^{18}$ to illustrate typical behavior.}
\end{figure} 

The dependence of the final value of $X_H$ on $G$ is shown in Figure 10 for different values of $f_{w-e}$, ranging from $10^{15}$ to $10^{23}$.  
The value of $f_{w-e}$ at which $X_H$ is significantly reduced decreases with increasing $G$, due to the tighter binding of diproton reducing its 
rate of photodestruction.  Figure 11 shows how the final value of the $^2$H abundance depends on $G$ for the same range of $f_{w-e}$. Clearly a 
major difference from standard BBN is in the amount of $^2$H that can be produced when the diproton is bound. To understand why consider the 
specific case $G = 1.2$. The binding energy of $^2$H is then about 10 MeV which means that it can be produced very early on in the big bang.  
Subsequent reactions reduce the $^2$H abundance by making $^3$He and $^4$He. Most of the $^4$He is produced very quickly (90\% is produced by $t 
= 8$ s). On the other hand, the binding energy of the diproton is relatively small, 0.7 MeV. Hence the temperature must drop to about $1.4 ~  
10^9$ K before significant production can begin. This occurs at $t = 100$ s. The diproton beta-decay life time is about $10^3$ s. Hence the 
diproton abundance increases during the first few thousand seconds, and then it decays to $^2$H. The temperature ($< 3 ~ 10^8$ K) is now too low 
for further reactions involving destruction of $^2$H. Hence, depending on the value of $f_{w-e}$, significant amounts of $^2$H can be produced.

\begin{figure}
\includegraphics[scale=0.5]{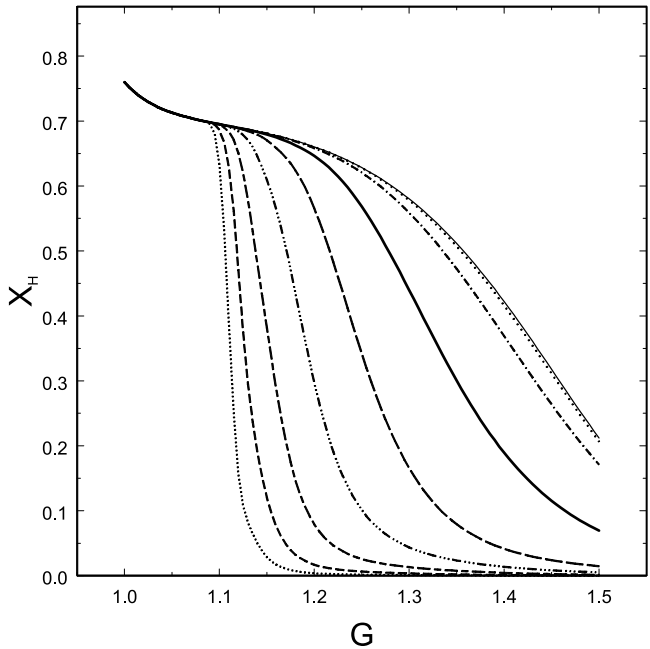}
\caption{\label{fig_10} Dependence of the final hydrogen abundance on $G$ for different values of $f_{w-e}$ ranging from $10^{15}$ at top to 
$10^{23}$ at bottom.}
\end{figure} 

Figure 12 shows the dependence of the final $^4$He abundance on $G$ for a range of $f_{w-e}$ values.  It can be seen that there are many 
combinations of $G$ and $f_{w-e}$ for which complete conversion to $^4$He does not occur.

\begin{figure}
\includegraphics[scale=0.5]{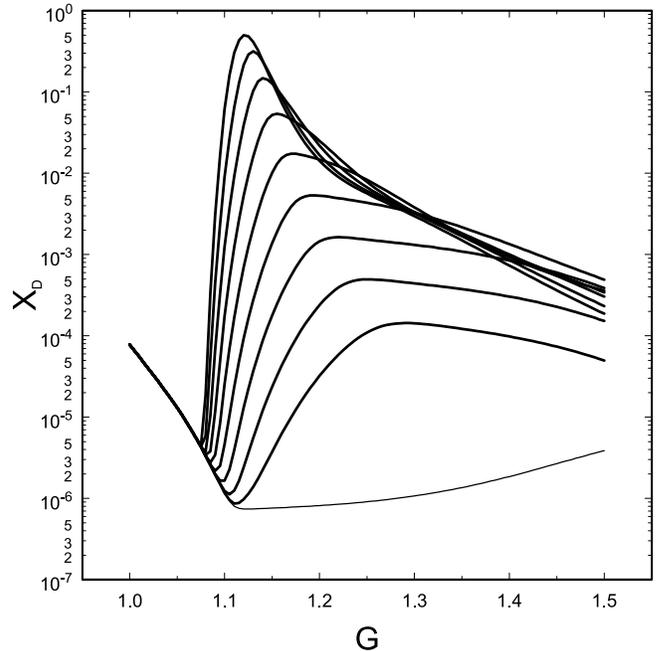}
\caption{\label{fig_11} Dependence of the final $^2$H abundance on $G$ for different values of $f_{w-e}$. The thin line is for $f_{w-e} = 0$. The 
thick lines are for $f_{w-e}$ ranging from $10^{15}$ (bottom) to $10^{23}$ (top).}
\end{figure} 

\begin{figure}
\includegraphics[scale=0.5]{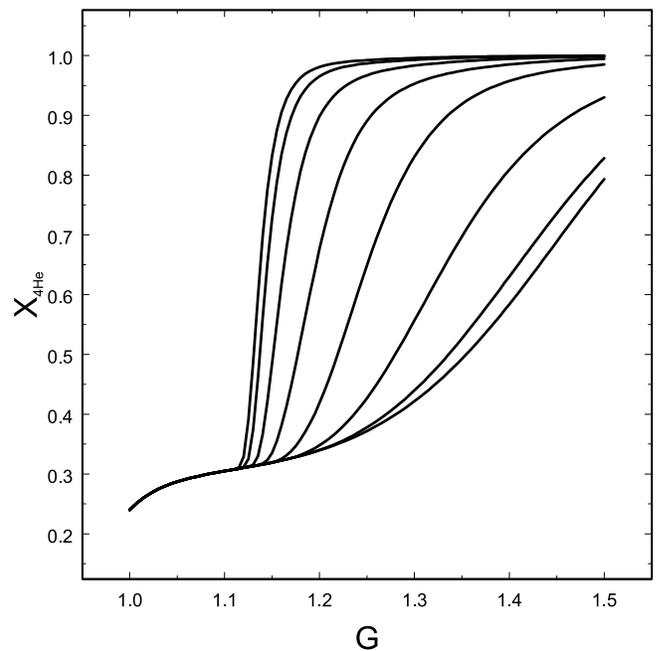}
\caption{\label{fig_12}Dependence of the final $^4$He abundance on $G$ for different values of $f_{w-e}$ ranging from $10^{16}$ (bottom) to
 $10^{23}$ (top). }
\end{figure} 
 
For $f_{w-e} = 10^{18}$  the final H abundance is greater than 10\% of the standard value for $G < 1.5$.  Hence significant amounts of H remain 
even when the strong force coupling constant is 50\% greater than the current value.  In general, the final H abundance is greater than 0.075 
provided  $f_{w-e} < 6 ~ 10^{15} / (G - 1.065)^6$.

We now consider inclusion of $pp + n \rightarrow d + p$ and $nn + p \rightarrow d + n$ as instantaneous reactions. In general these reactions 
have small effects on the final abundances, primarily because most of the neutrons have been depleted by the p + n reaction before the 
temperature has dropped sufficiently for diproton and dineutron production to occur.  A small amount of neutrons are produced during the second 
phase of nucleosynthesis by the $p+p \rightarrow pp + \gamma$, $pp \rightarrow d + e^+ + \nu$, $d + d \rightarrow t + p$, $t + t \rightarrow $$^4$$He 
+ n + n$ sequence of reactions. These neutrons can then react by $n + pp \rightarrow d + p$. The net result is a small increase in the final H 
abundance. Also since the $n + pp \rightarrow d + p$ reaction is assumed instantaneous, $^2$H is produced earlier than by diproton decay alone. 
Provided the temperature is high enough, this leads to a decrease in the final $^2$H abundance.

\begin{figure}
\includegraphics[scale=0.5]{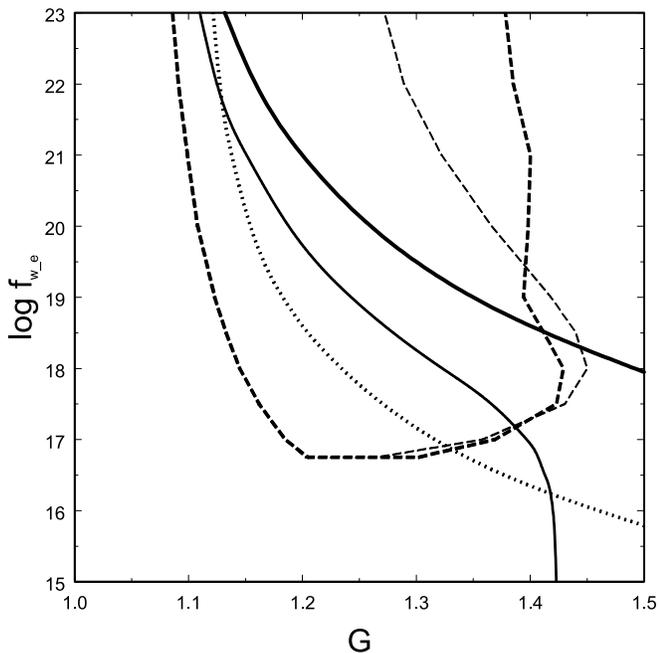}
\caption{\label{fig_13} The solid lines are contours on which the final H mass fraction is 0.075 (thick) and 0.375 (thin). The thin and thick 
broken lines are the contours on which the final $^2$H mass fraction is 0.001 with and without the pp + n and nn + p reactions, respectively. The 
dotted line is the second $^4$He production phase boundary.  Below and to the left of this line the second phase does not occur.}
\end{figure} 

Figure 13 summarizes the results of this section. The thicker of the solid lines is the contour on which the final H mass fraction is 0.075. The 
thinner solid line is the contour for final H mass fraction equal to 0.375, which is approximately half the standard BBN value. The broken lines 
are contours on which the final $^2$H mass fraction is 0.001. The thin solid line is the second $^4$He production phase boundary.  Below and to 
the left of this line the second phase does not occur.

\vspace{0.16666in}
\section{Conclusions}

We have addressed some aspects of the effects of larger than standard values for the strong force coupling constant on nucleosynthesis during the 
hot big bang. For relative strong charge $G > 1.065$, both the diproton and dineutron are bound. We have estimated the beta-decay time scales 
from the $ft$ factors for the analog nuclei $^{14}$O and $^{14}$C. Assuming that the rate of the reaction $p + p \rightarrow pp + \gamma$ can be 
parameterized by multiplying the rate of the reaction $p + p \rightarrow d + e^+ + \nu$ by a factor $f_{w-e}$, and that the rate of the reaction 
$n + n \rightarrow nn + \gamma$ is then related to that for the reaction $p + p \rightarrow pp + \gamma$ by neglecting the Coulomb repulsion, we 
find that significant amounts of H remain provided $f_{w-e} < 6 ~ 10^{15} / \left( G - 1.065 \right)^6$. By comparing similar reactions, we 
estimate that $f_{w-e} \sim 10^{18}$, which gives a corresponding limit of $G < 1.5$.  The primary reason for the survival of 
hydrogen is that the diproton and dineutron are always less tightly bound than the deuteron, which is a consequence of the spin-dependent part of 
the nuclear force. Photodestruction reactions prevent buildup of diprotons and dineutrons before the neutrons are depleted by deuteron formation. 
Diprotons can be formed once the temperature has dropped sufficiently. These diprotons are converted to deuterons mainly by beta decay with 
possibly a contribution from the $pp + n \rightarrow d + p$ reaction.  This can lead to much a larger $^2$H abundance than in the standard BBN.

Our main result is that the existence of bound diproton and dineutron nuclei does not necessarily lead to complete conversion of hydrogen to 
helium in the big bang. Instead there are parameter ranges for which significant amounts of hydrogen remain.  We estimate for reasonable values 
of the factor by which the
 $p + p \rightarrow pp + \gamma$ rate is enhanced relative to the $p + p \rightarrow d + e^+ + \nu$ rate, the final hydrogen abundance  is 
greater than 50\% of the standard BBN value for increases in the strong force coupling constant less than about 50\%. 
 Anthropic limits on the strong force strength from BBN are indeed weak.

\begin{acknowledgments}
We thank Stuart Pittel, David Seckel and Stephen Barr for enlightening discussions.  This research was supported in part by a grant from the 
Mount Cuba Astronomical Foundation.
\end{acknowledgments}

\bibliographystyle{apsrev}
\bibliography{BBN_anthropic}

\end{document}